\documentclass[aps,prd,notitlepage,nofootinbib,preprintnumbers,superscriptaddress,10pt]{revtex4-2}
\usepackage{here}
\usepackage{graphicx}
\usepackage{amsmath,amsthm,amssymb}
\usepackage{bm}
\usepackage{color}

\usepackage{amsfonts}
\usepackage{dcolumn}
\usepackage{hyperref}
\allowdisplaybreaks[1]
\usepackage{stackengine}

\newcommand{\be}{\begin{eqnarray}}
\newcommand{\ee}{\end{eqnarray}}

\newcommand{\bem}{\begin{bmatrix}}
\newcommand{\eem}{\end{bmatrix}}

\allowdisplaybreaks
\begin{document}
\title{Exact pp-wave solutions in shift-symmetric higher-order scalar-tensor theories}

\author{Masato Minamitsuji}
\affiliation{Faculty of Health Sciences, Butsuryo College of Osaka, Sakai 593-8328, Osaka, Japan}

\begin{abstract}
We investigate exact plane-fronted gravitational-wave (pp-wave) solutions within the framework of shift-symmetric quadratic-order higher-order scalar-tensor (HOST) theories. These solutions represent fully nonlinear radiative spacetimes that extend beyond the linearized approximation. We demonstrate that under the algebraic conditions on the coupling functions, the gravitational field equations reduce to a two-dimensional Laplace equation for the wave profile, recovering the structural form of vacuum general relativity. By adopting a scalar field ansatz that depends linearly on transverse coordinates and arbitrarily on the retarded null coordinate, we maintain a constant kinetic term of the scalar field. This configuration allows for a \emph{``stealth pp-wave''} solution, where a nontrivial scalar field profile coexists with the gravitational wave without backreacting on the spacetime geometry. We further show that these stealth configurations are fully compatible with the degeneracy conditions of class-Ia DHOST theories and satisfy current observational constraints. Finally, we examine the behavior of these solutions under disformal transformations, revealing that while the Brinkmann form is preserved, the stealth property is generically lost due to the mixing of scalar and tensor degrees of freedom. These results establish the robustness of pp-wave solutions in viable DHOST frameworks and highlight their utility for probing nonlinear effects in modified gravity.
\end{abstract}

\date{\today}

\maketitle

%%%%%%%%%%%%%%%%%%%%%%%%%%%%%%%%%%%%%%%%%%
\section{Introduction}
\label{sec1}

The detection of gravitational waves by the LIGO and Virgo Collaborations has opened a new observational window into strong-field gravity, offering opportunities to test the foundations of general relativity (GR) and its modifications \cite{LIGOScientific:2016aoc,Abbott:2016nmj}.  
Among the striking exact solutions of GR the so-called plane-fronted gravitational waves with parallel rays, or \emph{pp-waves}, represent exact radiative spacetimes propagating at the speed of light with planar wavefronts \cite{Penrose:1965rx,Sippel:1986if,Cianci:2015pba,Roche:2022bcz}.
Despite their mathematical simplicity, pp-wave spacetimes fully capture the nonlinear structure of GR and serve as powerful laboratories for probing fundamental properties of gravitational radiation.

From the theoretical viewpoint, scalar-tensor extensions of GR have received considerable attention as candidate theories of dark energy and modified gravity \cite{Fujii:2003pa,Horndeski,Deffayet:2009wt,Kobayashi:2012kh,Gleyzes:2014dya,Berti:2015itd}.  
In particular, degenerate higher-order scalar–tensor (DHOST) theories provide the most general class of scalar-tensor theories with higher-derivative interactions that remain free of Ostrogradsky instabilities \cite{Langlois:2015cwa,BenAchour:2016fzp,Crisostomi:2017lbg}.  
These theories cover a wide variety of models, including Horndeski and beyond-Horndeski theories \cite{Horndeski,Deffayet:2009wt,Kobayashi:2012kh,Gleyzes:2014dya}, and are characterized by a set of degeneracy conditions ensuring the absence of extra ghostly degrees of freedom.  
Observational constraints from the coincidence of the speed of gravitational waves and light, inferred from the binary neutron star merger GW170817 \cite{TheLIGOScientific:2017qsa,LIGOScientific:2017vwq,LIGOScientific:2017zic,Monitor:2017mdv}, further restrict viable subclasses of DHOST theories \cite{Baker:2017hug,Creminelli:2017sry,Ezquiaga:2017ekz,Langlois:2017dyl,Kase:2018aps}.
Furthermore, no perturbative or nonperturbative decay of gravitational waves into dark energy imposes further constraints on the theories \cite{Creminelli:2018xsv,Creminelli:2019nok}.

We would like to ask whether the exact pp-wave solutions of GR continue to exist within the framework of DHOST theories.
In this work, we focus on quadratic-order shift-symmetric higher-order scalar-tensor (HOST) theories without imposing the degeneracy conditions. 
Such theories are constructed from quadratic combinations of the second-order derivatives of the scalar field, and include, in addition, a generalized kinetic term, a generalized cubic Galileon term, and the nonminimal coupling to the Ricci tensor.
We first examine whether the pp-wave solutions remain valid solutions within quadratic-order shift-symmetric HOST theories. 
We then investigate their compatibility with the degeneracy conditions that characterize class Ia quadratic-order DHOST theories, as well as with additional constraints arising from observational viability, such as the consistency with gravitational-wave propagation.

Starting from a generalized pp-wave ansatz for the metric and a scalar field profile depending on the null and transverse coordinates, we derive the corresponding Euler-Lagrange equations and then investigate whether these Euler-Lagrange equations are satisfied for the pp-wave metric and the form of the scalar field compatible with the shift symmetry. 
If such solutions exist, they could be useful for investigating the propagation of gravitational and scalar waves in modified gravity, and clarify the interplay between nonlinear gravitational radiation and scalar degrees of freedom. 
The possible existence of {\it stealth} scalar configurations, which are nontrivial scalar field profiles that do not backreact on the background geometry and are originally investigated for static or stationary black hole solutions \cite{Mukohyama:2005rw,Babichev:2013cya,Babichev:2016rlq,Motohashi:2019sen,Takahashi:2019oxz,Motohashi:2019ymr,Takahashi:2020hso,DeFelice:2022qaz}, offers an intriguing mechanism by which scalar degrees of freedom may coexist with gravitational waves without altering their spacetime structure.

Previous studies of gravitational waves in modified gravity theories have mostly focused on the linearized regime, considering small perturbations around Minkowski or cosmological backgrounds \cite{Barack:2018yly}. By contrast, the fate of exact nonlinear radiative solution, such as pp-wave configurations, have not been unexplored much in HOST theories, unless one performs a fully nonlinear analysis. 
Understanding these solutions is crucial for assessing the robustness of gravitational-wave propagation and their potential interactions with additional scalar degrees of freedom inherent in modified gravity frameworks.  

Exact solutions provide genuinely nonperturbative information about the dynamics of the theory that cannot be captured by linearized analyses, revealing effects that may only appear at strong-field or highly nonlinear regimes~\cite{Babichev:2012qs,BenAchour:2024zzk,BenAchour:2024tqt}.
Such solutions serve as theoretical laboratories for systematically testing fundamental aspects of the propagation of scalar and tensor degrees of freedom, their mutual interactions, and the possible existence of stealth configurations where scalar fields coexist with gravitational waves without backreacting on the spacetime geometry. 
They help to clarify the consistency of the theory with the fundamental principle that gravitational radiation should propagate without pathological behavior even in the high-frequency limit or in the presence of strong-field effects.

We will find that under only a few mild conditions on the coupling functions the resulting Euler-Lagrange equations reduce to a two-dimensional Laplace equation for the wave profile $H(u,x_1,x_2)$, precisely matching the condition required for the existence of exact pp-wave solutions in vacuum GR. 
The scalar field is then chosen to have a linear dependence on the transverse coordinates, supplemented by an arbitrary function of the retarded time $u$, which leads to a constant kinetic term $X = X_0$, where $X$ is the kinetic term of the scalar field (see Sec. \ref{sec2}), and is fully compatible with the shift symmetry of the theory.  
Imposing consistency with the Euler-Lagrange equations constrains only simple algebraic conditions on the potential function $F_0(X)$ evaluated at $X = X_0$. When these conditions are satisfied, the scalar field propagates alongside the gravitational wave without affecting the spacetime geometry—a configuration we interpret as a \emph{stealth pp-wave} solution. 
We further demonstrate that such pp-wave solutions of GR persist in the viable subclasses of quadratic-order shift-symmetric DHOST theories, accompanied by a scalar field that does not backreact on the metric. 
This result demonstrates the robustness of pp-waves as exact solutions beyond GR, reinforces their utility as theoretical probes for exploring modified gravity, and opens a pathway for future studies of nonlinear gravitational radiation in higher-order scalar-tensor frameworks, including potential applications to observational signatures and fundamental consistency tests.

\section{The model}%%%%%%%%%%%%%%%%%%%%%%%%%%%%%%%%%%%%%%%%%%%%%%%%%%%%%%%%
\label{sec2}

We consider the quadratic-order HOST theories introduced in~\cite{Langlois:2015cwa}, whose Lagrangian contains quadratic contractions of the second derivatives of the scalar field. The action of the HOST theories is given by
\begin{equation}
\label{qdaction}
S = \int d^4x \sqrt{-g}\,
\left[
F_0(\phi,X)
+ F_1(\phi,X)\, \Box\phi
+ F_2(\phi,X)\, R
+ \sum_{I=1}^5 A_I(\phi,X)\, L_I^{(2)}
\right],
\end{equation}
where $g_{\mu\nu}$ is the spacetime metric, $g := \det(g_{\mu\nu})$ denotes its determinant, and $R$ is the Ricci scalar constructed from $g_{\mu\nu}$. The scalar degree of freedom is represented by $\phi$, with kinetic term $X := \phi_\mu \phi^\mu$, where $\phi_\mu := \nabla_\mu \phi$ denotes the covariant derivative of the scalar field with respect to the spacetime coordinate $x^\mu$.

The five quadratic Lagrangians $L_I^{(2)}$ appearing in Eq.~\eqref{qdaction} are given by
\begin{equation}
L_1^{(2)} = \phi^{\mu\nu}\phi_{\mu\nu}, \qquad
L_2^{(2)} = (\Box\phi)^2, \qquad
L_3^{(2)} = \phi^\mu \phi_{\mu\nu} \phi^\nu \Box\phi, \qquad
L_4^{(2)} = \phi^\mu \phi_{\mu\nu} \phi^{\nu\lambda} \phi_\lambda, \qquad
L_5^{(2)} = (\phi^\mu \phi_{\mu\nu} \phi^\nu)^2,
\end{equation}
where $\phi_{\mu\nu} := \nabla_\nu \nabla_\mu \phi$ denotes the second covariant derivative of $\phi$, and $\Box\phi := g^{\mu\nu}\phi_{\mu\nu}$ is the d'Alembertian. The functions $F_i(\phi,X)$ ($i=0,1,2$) and $A_I(\phi,X)$ ($I=1,\ldots,5$) are arbitrary scalar functions of their arguments, thereby parametrizing a very general class of higher-order scalar-tensor interactions.

The quadratic-order DHOST theories arise as special subclasses of Eq.~\eqref{qdaction}, identified by degeneracy conditions that eliminate unwanted Ostrogradsky instabilities.
Within this large family, a particularly relevant class is provided by the class-Ia DHOST theories, which ensure that the propagation speed of tensor modes ($c_t$) matches the speed of light ($c$). 
Such theories are motivated by the stringent observational bounds on the speed of gravitational waves from the binary neutron star merger GW170817~\cite{Langlois:2017dyl}. The degeneracy conditions specifying this subclass read
\be \label{ctc}
A_1=A_2=0, \quad
A_4 = \frac{1}{8F_2}\left[48F_{2X}^2 - 8(F_2 - X F_{2X}) A_3 - X^2 A_3^2 \right], \quad
A_5 = \frac{1}{2F_2}(4F_{2X}+X A_3)A_3 ,
\ee
where $A_3$ remains a free function of $(\phi,X)$.
An additional physical requirement further constrains this subclass. In particular, it has been argued that, in the presence of a nonvanishing $A_3$, gravitational waves would generically suffer a perturbative or even nonperturbative decay into fluctuations of the scalar sector, interpreted as dark energy~\cite{Creminelli:2018xsv,Creminelli:2019nok}. To forbid such an instability, one must impose
\be \label{nodecay}
A_3 = 0 .
\ee
Substituting this condition back into Eq.~\eqref{ctc}, the degeneracy conditions simplify to
\be \label{ctc2}
A_1=A_2=0, \quad
A_4 = 6\frac{F_{2X}^2}{F_2}, \quad
A_5 = 0 .
\ee
This form defines the so-called healthy class-Ia DHOST theories, compatible with both theoretical stability and observational constraints on gravitational-wave propagation.

\medskip

Finally, we impose a global shift symmetry $\phi \to \phi + c$, with $c$ constant. Under this assumption, all functions in the action \eqref{qdaction} depend solely on the kinetic term $X$, thereby eliminating explicit $\phi$ dependence. The shift symmetry not only reduces the parameter space of admissible theories but also plays an important role in protecting the scalar sector from large quantum corrections. Consequently, the setup described above represents one of the most theoretically consistent and observationally viable corners of the quadratic HOST/DHOST landscape.

\section{Plane-fronted gravitational-wave solutions}

\subsection{pp-wave solutions in GR}

A particularly important class of exact solutions to the Einstein equations in GR is the family of plane-fronted gravitational waves with parallel rays, or \emph{pp-waves} for short.  
These spacetimes describe gravitational radiation propagating at the speed of light with planar wavefronts, and they fully incorporate the nonlinear structure of GR while retaining a high degree of mathematical simplicity.

The general pp-wave metric can be written in Brinkmann coordinates $(u,v,x_1,x_2)$ as
\begin{align}
\label{pp_wave}
ds^2
=
H(u,x_1,x_2)\,du^2
+
2\,du\,dv
+
dx_1^2
+
dx_2^2,
\end{align}
where $u$ is the retarded null coordinate, representing the phase of the wave,
and $v$ is the affine parameter along the null geodesics associated with the covariantly constant null vector field $\ell^\mu = \left(\partial_v\right)^\mu$. It should be noted that $v$ is not a null coordinate, since the hypersurfaces $v=\text{const}$ are not null.
 $(x_1,x_2)$ are Cartesian coordinates on the two-dimensional transverse space orthogonal to the propagation direction.  
$H(u,x_1,x_2)$ is an arbitrary function specifying the wave profile.
The wave propagates in the direction of increasing $u$, with each surface $u=\mathrm{const}$ representing a wavefront of constant phase. These are null hypersurfaces that carry the entire curvature content of the spacetime.
The pp-wave spacetimes are members of the Kundt class of metrics and are characterized by the existence of a covariantly constant null vector field $\ell^\mu = (\partial_v)^\mu$, satisfying
\begin{align}
\nabla_\nu \ell^\mu = 0, \qquad \ell^\mu \ell_\mu = 0.
\label{KUNDT_COND}
\end{align}
The vector $\ell^\mu$ is tangent to the null geodesics that generate the wavefronts and is parallel-transported everywhere in spacetime.  
This property reflects the fact that the wave rays remain parallel during propagation, hence the parallel rays.
The only nontrivial components of the Ricci tensor for the metric \eqref{pp_wave} are
\begin{align}
R_{uu} = -\frac{1}{2} \left( \partial_{x_1}^2 H + \partial_{x_2}^2 H \right),
\end{align}
with all other components vanishing.  
Thus, the vacuum Einstein equations in GR $G_{\mu\nu} = 0$ reduce to the two-dimensional Laplace equation on the transverse space:
\begin{align}
\label{pp_wave_laplace}
\partial_{x_1}^2 H
+
\partial_{x_2}^2 H
=0.
\end{align}
Any $H(u,x_1,x_2)$ satisfying Eq.~\eqref{pp_wave_laplace} yields an exact Ricci-flat pp-wave spacetime.
We note that Eq.~\eqref{pp_wave_laplace} is linear in $x_1$ and $x_2$, even though the underlying Einstein equations  in GR are fully nonlinear in the metric components.  
This means that the transverse spatial profile of the wave can be obtained by superposition of harmonic functions, but the dependence on $u$ can be completely arbitrary, allowing for highly nontrivial temporal behavior.
A particularly simple and physically relevant subclass of pp-wave metrics is obtained by choosing $H$ to be quadratic in the transverse coordinates,
\begin{align}
\label{pc}
H(u,x_1,x_2) = A(u)\,(x_1^2 - x_2^2) + 2\,B(u)\,x_1 x_2.
\end{align}
Here, $A(u)$ and $B(u)$ are arbitrary functions of $u$ representing the wave profile in the two independent polarization states of the gravitational waves.
The term proportional to $A(u)$ corresponds to the plus ($+$) polarization, producing tidal distortions along the $x_1$ and $x_2$ axes in opposite directions. The term proportional to $B(u)$ corresponds to the cross ($\times$) polarization, producing tidal distortions along axes rotated by $45^\circ$ with respect to $(x_1,x_2)$.
These polarizations are directly analogous to the $+$ and $\times$ modes appearing in the linearized theory of gravitational waves in flat spacetime.  
However, in the pp-wave metric \eqref{pc}, the gravitational field is exact and includes all nonlinear effects of GR.  
Thus, Eq.~\eqref{pc} may be regarded as a nonlinear generalization of the linearized plane-wave solution.
In the weak-field approximation, one considers a small perturbation $h_{\mu\nu}$ around Minkowski spacetime.  
For a wave propagating along the $+z$ direction, the $+$ and $\times$ modes appear as independent components of $h_{ij}$ in the transverse-traceless gauge.  
If $A(u)$ and $B(u)$ are small, the pp-wave metric \eqref{pc} reproduces exactly this linearized form.  
For large amplitudes, however, the solution remains valid due to the exact nature of the Brinkmann form.

In many physical situations, gravitational waves are not monochromatic but are instead localized bursts or pulses, produced by transient astrophysical events such as black hole mergers or supernova explosions.  
A pp-wave solution modeling such a burst can be written as
\begin{align}
H(u,x_1,x_2) = f(x_1,x_2)\,\mathrm{sech}^2\!\left( \frac{u}{u_0} \right),
\end{align}
where $f(x_1,x_2)$ encodes the transverse spatial structure of the wave and satisfies the harmonic condition
    \begin{align}
    \partial_{x_1}^2 f + \partial_{x_2}^2 f = 0,
    \end{align}
ensuring the equations of motion are satisfied in vacuum GR. The factor $\mathrm{sech}^2(u/u_0)$ localizes the wave in the $u$ direction, producing a short-duration pulse of width $\sim u_0$.
Such a profile vanishes exponentially as $|u| \to \infty$, describing a gravitational wave that is nonzero only during a finite interval of retarded time. The $\mathrm{sech}^2$ envelope could model a wave packet with a single dominant burst in the $u$ direction.
The harmonic condition for $f(x_1,x_2)$ ensures that the curvature is confined to the null hypersurfaces of constant $u$, preserving the defining feature of pp-waves.

The curvature of a pp-wave spacetime is entirely encoded in the Riemann tensor components
\begin{align}
R_{uiuj} = -\frac{1}{2} \,\partial_{x_i} \partial_{x_j} H(u,x_1,x_2),
\end{align}
where $i,j = 1,2$ label the transverse coordinates.  
This has direct physical significance. 
The geodesic deviation equation for two nearby freely falling test particles with separation vector $\xi^i$ in the transverse plane reads
\begin{align}
\frac{d^2 \xi^i}{du^2} = - R^i{}_{uju}\,\xi^j.
\end{align}
For the plane-wave profile \eqref{pc}, this becomes
\begin{align}
\frac{d^2}{du^2}
\begin{pmatrix}
\xi^1 \\ \xi^2
\end{pmatrix}
=
-
\begin{pmatrix}
A(u) & B(u) \\
B(u) & -A(u)
\end{pmatrix}
\begin{pmatrix}
\xi^1 \\ \xi^2
\end{pmatrix}.
\end{align}
This clearly shows that $A(u)$ and $B(u)$ govern the tidal stretching and squeezing experienced by particles in the transverse plane, providing a direct operational meaning to the gravitational-wave polarizations in the exact theory.

Conserved quantities in radiative spacetimes are naturally defined at future
null infinity within the Bondi-Sachs framework, where the Bondi mass decreases
according to the standard mass-loss formula involving the news tensor $N_{AB}$.
Pure pp-wave geometries, however, are not asymptotically flat in the spherical
sense and instead admit a planar-type null infinity generated by the null
Killing vector $\ell^\mu=\partial_v$.  In this setting the Bondi-type mass
vanishes identically, while the analog of the news tensor in the transverse
plane is nonzero whenever $\partial_u H\neq 0$, reflecting the radiative nature
of the wave.  Although the mass is zero, conserved momentum and angular-momentum
fluxes arise from Killing symmetries of the transverse plane, and in vacuum
pp-waves these quantities are purely geometrical.

\subsection{pp-wave solutions in HOST theories}

In order to investigate whether the vacuum pp-wave solutions of GR also persist
as exact solutions within the framework of HOST theories, we begin with the most
general metric ansatz compatible with the existence of a null coordinate and a
two-dimensional Euclidean transverse space.
From a purely geometric perspective, pp-waves are characterized by the existence of a covariantly constant null vector field $\ell$ satisfying Eq. \eqref{KUNDT_COND}, i.e., $\nabla_\nu \ell_\mu = 0$. This condition indicates, via the Frobenius theorem, that $\ell_\mu$ is hypersurface orthogonal ($\ell_{[\mu} \nabla_\nu \ell_{\rho]} = 0$), which allows us to label the corresponding null hypersurfaces with a null coordinate $u$. Furthermore, because $\ell$ is geodesic by definition, we can choose the affine parameter $v$ along the null rays as a second coordinate such that $\ell^\mu =( \partial_v)^\mu$. Combined with the two transverse coordinates $(x_1, x_2)$ of a Euclidean two-space, this geometric structure strongly restricts the form of the metric and its coordinate dependence from the outset.
Importantly, we do not assume that a pp-wave solves the HOST field equations; the restrictions imposed below follow solely from coordinate freedom and symmetry considerations.

We introduce a null coordinate $u$ and a second coordinate $v$ adapted to the
corresponding null direction, together with two transverse coordinates
$(x_1,x_2)$ parametrizing a two-dimensional Euclidean space.  Starting from a
general metric containing all components
\[
g_{uu},\ g_{uv},\ g_{vv},\ g_{ui},\ g_{vi},\ g_{ij},
\qquad i,j=1,2,
\]
we impose only the following geometric requirements:

\begin{description}

\item[
(i)]The metric components are independent of $v$, reflecting the fact that the covariantly constant null vector $\ell$ is, in particular, a Killing vector field.

\item[(ii)] 
The transverse sector admits a two-dimensional Euclidean symmetry, so that
$g_{ij}$ can be brought to a conformally flat form.

\end{description}

Under these assumptions, the remaining coordinate (gauge) freedom consists of
transformations of the type
\begin{align}
u \rightarrow u'(u), \qquad
v \rightarrow v' = v + \alpha(u,x_1,x_2), \qquad
x_i \rightarrow x'_i = {\cal R}_{ij}(u)\,x_j + b_i(u),
\end{align}
where ${\cal R}_{ij}(u)$ is a $u$-dependent rotation matrix and $b_i(u)$ are
arbitrary functions.  These transformations act on the metric components as
follows:

\begin{enumerate}

\item
The shift $v \rightarrow v + \alpha(u,x)$ removes the $dv\,dx_i$ terms and
  allows the coefficient of $du\,dv$ to be normalized.

\item
The $u$-dependent transverse shifts $x_i \rightarrow x_i + b_i(u)$ eliminate
  the $du\,dx_i$ components locally.
  It should be noted that while these terms can be removed by local coordinate transformations, they may globally encode gyratonic effects related to the intrinsic angular momentum of the pp-wave source~\cite{Frolov:2005in,Podolsky:2014lpa}. In this work, we focus on solutions where these terms are absent.

\item
The $u$-dependent rotations ${\cal R}_{ij}(u)$ bring the transverse metric
  $g_{ij}$ to a conformally flat form.

\end{enumerate}

See Appendix \ref{app_gauge}
for details of the gauge fixing.
Since these simplifications rely solely on coordinate freedom and not on the
field equations, the metric can always be written, without loss of generality,
as
\begin{eqnarray}
\label{pp_wave_general}
ds^2
&=&
H(u,x_1,x_2)\, du^2
+ I(u,x_1,x_2)\, dv^2
+ 2\,V(u,x_1,x_2)\, du\, dv
+ W(u,x_1,x_2)\,(dx_1^2 + dx_2^2),
\nonumber\\
\phi &=& \phi(u,x_1,x_2),
\end{eqnarray}
where the four functions $H$, $I$, $V$, and $W$ encode all physically distinct
degrees of freedom compatible with the chosen coordinate system.  The scalar
field $\phi$ is allowed to depend on all coordinates except $v$.

Before substituting the ansatz \eqref{pp_wave_general} into the field equations, we clarify the relationship between this metric form and the pp-wave class. The fundamental property of a pp-wave is the existence of a covariantly constant null vector field $\ell^\mu$ satisfying the conditions in Eq.~\eqref{KUNDT_COND}. 
For the metric \eqref{pp_wave_general}, the requirement that $\ell^\mu = (\partial_v)^\mu$ be a null and covariantly constant vector field $(\nabla_\nu \ell^\mu = 0)$ directly implies $I=0$ and the condition $\nabla_i \ell_u = 0$ restricts $V$ to be a function of $u$ only, which can then be normalized to $V=1$ by using the freedom to choose $v$ as an affine parameter.
While these conditions ensure that the null rays are parallel, the defining property of the pp-wave class further requires the transverse space to be flat, which is guaranteed by taking $W$ to be a constant, normalized to $W=1$.
Thus, the conditions
\begin{align}
I(u,x_1,x_2) = 0, \qquad
V(u,x_1,x_2) = 1, \qquad
W(u,x_1,x_2) = 1,
\label{pp_wave_geom_cond}
\end{align}
are not merely a convenient gauge choice but constitute the essential geometric requirements to specialize the general Kundt-type metric \eqref{pp_wave_general} to the canonical pp-wave form
\begin{align}
ds^2 = H(u,x_1,x_2)\, du^2 + 2\,du\,dv + dx_1^2 + dx_2^2 .
\label{pp_wave_host}
\end{align}
In the following, we investigate whether this specific geometric class admits solutions within the framework of HOST theories.
After deriving the Euler-Lagrange equation for the scalar field, we adopt the
ansatz
\begin{align}
\label{scalar_pp}
\phi(u,x_1,x_2) = \phi_0(u) + q_1 x_1 + q_2 x_2 ,
\end{align}
where $q_1$ and $q_2$ are nonzero constants and $\phi_0(u)$ is an arbitrary
function of $u$.  For the metric \eqref{pp_wave_host} and scalar field
\eqref{scalar_pp}, the scalar kinetic term takes the constant value
\begin{align}
X = q_1^2 + q_2^2 \equiv X_0 > 0,
\label{x0}
\end{align}
which follows from the linear dependence of the scalar field on the transverse
coordinates and is consistent with the shift symmetry
$\phi \rightarrow \phi + \mathrm{const}$.

The field equations imply that the function
$H(u,x_1,x_2)$ must satisfy the two-dimensional harmonic equation
\begin{align}
\partial_{x_1}^2 H + \partial_{x_2}^2 H = 0,
\label{pp_wave_laplace_host}
\end{align}
identical to the Ricci-flat condition for pp-waves in vacuum GR.  The general
solution of Eq.~\eqref{pp_wave_laplace_host} can be expressed in terms of
harmonic polynomials or as superpositions of plane-wave modes in the transverse
coordinates.

The remaining Euler-Lagrange equations reduce to the algebraic compatibility
conditions
\begin{align}
\label{pp_cond}
F_0(X_0) = 0, \qquad
F_{0X}(X_0) = 0, \qquad
A_1(X_0) = 0,
\end{align}
where $F_{0X} \equiv dF_0/dX$.  These relations ensure that no effective
cosmological constant or tadpole term is generated.  They are also compatible
with the degeneracy condition \eqref{ctc} for class~Ia DHOST theories, which
removes Ostrogradsky ghosts, and with the no-decay condition \eqref{nodecay}.
One may alternatively impose the condition
\begin{align}
\label{s_branch}
F_2(X_0)=0 ,
\end{align}
under which an arbitrary profile $H(u,x_1,x_2)$---not necessarily satisfying
the Laplace equation \eqref{pp_wave_laplace_host}---seems to solve the field
equations.  This suggests the presence of a ``second branch'' of solutions.
However, the condition $F_2(X_0)=0$ corresponds to a vanishing gravitational
constant and is therefore physically unacceptable.  Moreover, once the Class-Ia
DHOST degeneracy conditions \eqref{ctc} are imposed, substituting
\eqref{s_branch} renders the denominators in \eqref{ctc} singular, indicating
that this branch is mathematically incompatible with the class-Ia structure.
Hence the apparent freedom to choose $H$ arbitrarily disappears, and the branch
\eqref{s_branch} cannot be realized within class-Ia DHOST theories.  The only
admissible solutions are those for which $H$ satisfies the Laplace equation
\eqref{pp_wave_laplace_host}.

The resulting solution represents a scalar field that propagates together with a gravitational pp-wave while remaining completely hidden from the gravitational dynamics.  
The term ``stealth'' refers to the remarkable situation in which the scalar field is fully nontrivial, with both a wave component depending on the
null coordinate $u$ and constant gradients in the transverse directions, yet its
energy-momentum tensor vanishes once the field equations are imposed. 
This vanishing is not a consequence of the scalar field being trivial.  
It arises
from a delicate cancellation among the various contributions that appear in HOST
theories, including terms involving first and second derivatives of the scalar
field.  
The cancellation ensures that the scalar field does not act as a source
for the geometry even though it carries nonzero gradients and possesses genuine
dynamical content.

Because the stress tensor disappears on shell, the geometry is identical to that
of a vacuum pp-wave in general relativity.  The curvature is determined entirely
by the harmonic function $H(u,x_1,x_2)$, and the characteristic structure of
pp-waves is preserved.  The null vector $\partial_v$ remains covariantly
constant, the Weyl tensor keeps its type N form, and the tidal forces
experienced by test particles are exactly those of a pure gravitational wave.
No observable quantity in the gravitational sector reveals the presence of the
scalar field.  From the viewpoint of GR alone, the spacetime
would be interpreted as a standard vacuum pp-wave without any additional degrees
of freedom.

Nevertheless, the scalar field  has a rich internal structure.  
The function
$\phi_0(u)$ can describe an arbitrary wave packet moving along the same null
direction as the gravitational wave.  The constants $q_1$ and $q_2$ generate a
uniform gradient across the transverse plane.  These components coexist without
disturbing the geometry.  The scalar field evolves on the pp-wave background as
if it were a test field, even though it is part of the fundamental gravitational
theory.  Both the gravitational wave and the scalar wave propagate at the speed
of light along the same null direction, yet they do not exchange energy.  The
absence of energy transfer follows from the stealth property.  The scalar field
has no gravitational imprint, and the gravitational wave does not influence the
scalar dynamics beyond providing a fixed background.

This situation raises an interesting question regarding observability.  A purely
gravitational probe would never detect the scalar field because all curvature
quantities coincide with those of a vacuum solution.  However, in a broader
theoretical context, the scalar field might couple to matter or to additional
fields.  If such couplings exist, the scalar field could leave observable
signatures that are not gravitational in nature.  
The stealth property therefore
does not imply absolute invisibility. 
It guarantees only that the scalar field
is invisible to the gravitational equations of motion.
The coexistence of a nontrivial scalar field with a geometry that behaves
exactly as a vacuum pp-wave highlights the special structure of HOST theories.
The cancellation required for stealth behavior is highly restrictive.  It
depends on the specific form of the Lagrangian and on the precise alignment
between the scalar field configuration and the null geometry of the pp-wave.
When these conditions are satisfied, the scalar field effectively rides on the
gravitational wave without disturbing it.  The result is a genuinely nontrivial
configuration in which two independent wave phenomena share the same spacetime
while remaining completely decoupled at the level of gravitational dynamics.

\section{Disformal and conformal transformations of the stealth pp-wave solutions}

Assuming that the pp-wave metric \eqref{pp_wave_host} together with the scalar
field \eqref{scalar_pp} provides a solution in a class of HOST theories, we now
examine how this solution transforms under a pure disformal transformation,
\begin{equation}
\label{disformal}
\bar g_{\mu\nu} = g_{\mu\nu} + B(X)\,\phi_\mu \phi_\nu,
\end{equation}
where $B$ is a function of the kinetic term
\begin{equation}
X = g^{\mu\nu}\phi_\mu \phi_\nu.
\end{equation}
Quadratic-order HOST theories are closed under such transformations.
Applying
\eqref{disformal} to a theory in this class generates another theory of the same
class; hence the transformation acts as a map between different quadratic-order
HOST theories without spoiling their structure.

The inverse metric associated with \eqref{disformal} is
\begin{equation}
\bar g^{\mu\nu}
= g^{\mu\nu}
- \frac{B}{1+BX}\,\phi^\mu \phi^\nu,
\end{equation}
so the map is invertible provided $1+BX\neq 0$.  The scalar kinetic term
transforms as
\begin{equation}
\bar X = \frac{X}{1+BX},
\end{equation}
which remains finite under the same nondegeneracy condition.

Applying \eqref{disformal} to the stealth pp-wave solution
\eqref{pp_wave_host}-\eqref{x0}, the transformed metric
takes the form
\begin{align}
\bar g_{\mu\nu} dx^\mu dx^\nu
&=
\Bigl(H + B\,\phi_0'(u)^2\Bigr) du^2
+ 2\,du\,dv
+ B q_1 \phi_0'(u)\, du\, dx_1
+ B q_2 \phi_0'(u)\, du\, dx_2
\nonumber\\
&\quad
+ (1 + B q_1^2) dx_1^2
+ (1 + B q_2^2) dx_2^2
+ 2 B q_1 q_2\, dx_1 dx_2.
\end{align}
Since $X=X_0$ for the stealth pp-wave, the function $B$ becomes a constant.
Using the coordinate redefinitions
\begin{align}
dx_1 &= -\frac{q_1 B \phi_0'(u)}{1 + B(q_1^2+q_2^2)}\,du
+ \frac{1}{\sqrt{1 + B q_1^2}}\, d\bar x_1
- \frac{B q_1 q_2}{\sqrt{(1 + B q_1^2)(1 + B(q_1^2+q_2^2))}}\, d\bar x_2, \\
dx_2 &= -\frac{q_2 B \phi_0'(u)}{1 + B(q_1^2+q_2^2)}\,du
+ \sqrt{\frac{1 + B q_1^2}{1 + B(q_1^2+q_2^2)}}\, d\bar x_2,
\end{align}
the metric reduces to the canonical pp-wave form
\begin{equation}
\label{pp_disformal}
\bar g_{\mu\nu} dx^\mu dx^\nu
= \bar H(u,\bar x_1,\bar x_2)\, du^2 + 2\, du\, d v + d\bar x_1^2 + d\bar x_2^2,
\end{equation}
with the shifted profile
\begin{equation}
\bar H = H + \frac{B\,\phi_0'(u)^2}{1 + B(q_1^2 + q_2^2)}.
\end{equation}
The kinetic term in the barred frame becomes
\begin{equation}
\bar X
= \frac{X_0}{1+BX_0}
= \frac{q_1^2+q_2^2}{1+B(q_1^2+q_2^2)}.
\end{equation}
Thus the disformal transformation maps a stealth pp-wave solution into another
solution with the same Brinkmann form but with a shifted profile $\bar H$ and a
rescaled kinetic term $\bar X$.
However, the transformed profile does not, in general, satisfy the pp-wave
equation.  Indeed, one finds
\begin{equation}
\label{disformal_laplace}
\partial_{\bar x_1}^2 \bar H
+
\partial_{\bar x_2}^2 \bar H
=
\frac{1}{1+B(q_1^2+q_2^2)}
\left[
(1+Bq_2^2)\,\partial_{x_1}^2 H
+
(1+Bq_1^2)\,\partial_{x_2}^2 H
-
2Bq_1q_2\,\partial_{x_1}\partial_{x_2}H
\right],
\end{equation}
which does not vanish even if $H$ satisfies the Laplace equation
\eqref{pp_wave_laplace_host} in the original frame.  Therefore, although the
transformed metric \eqref{pp_disformal} retains the canonical Brinkmann form, it
does not generically describe a pure pp-wave solution.  The disformal map mixes
the propagation of the gravitational and scalar waves, and the stealth property
is lost.  The only exception occurs when $q_1=q_2=0$, in which case the
transverse sector remains isotropic and the Laplace equation is preserved.
However, this corresponds to a vanishing kinetic term $X_0=0$, for which
perturbations are strongly coupled and the configuration is not physically
viable.

In summary, disformal transformations preserve the Brinkmann form of the metric
but do not preserve the pp-wave condition.  A stealth pp-wave solution in one
frame is mapped to a configuration in which the gravitational and scalar sectors
are nontrivially coupled, and the pp-wave character is lost.

\medskip

The construction naturally extends to the combined conformal-disformal
transformation
\begin{equation}
\label{disformal2}
\bar g_{\mu\nu} = A(X)\, g_{\mu\nu} + B(X)\,\phi_\mu \phi_\nu,
\end{equation}
where both $A$ and $B$ are arbitrary functions of the kinetic term $X$.  Since
quadratic-order HOST theories are closed under such transformations, a pp-wave
solution is mapped into another configuration of the same class, with its
profile $\bar H$ modified according to the values of $A(X_0)$ and $B(X_0)$.
This behavior is fully consistent with the closure of quadratic-order HOST
theories under disformal maps: the transformation acts as a reparametrization of
the metric that preserves the null structure $du\,dv$ characteristic of pp-wave
geometries, while shifting the profile $H$ and rescaling the kinetic term $X$.

Geometrically, the vector field $\phi_\mu$ defining the disformal deformation
selects a preferred direction in spacetime and thereby breaks the isotropy of
the transverse $(x_1,x_2)$ plane.  Consequently, the Laplace operator governing
the pp-wave profile is modified as in Eq.~\eqref{disformal_laplace}, and the
condition $\partial_{x_1}^2H+\partial_{x_2}^2H=0$ is no longer preserved.  This
does not signal any inconsistency; rather, it reflects the fact that the scalar
and tensor perturbations become dynamically coupled in the disformally related
frame.  A stealth configuration, in which the scalar field does not backreact on
the geometry, is thus mapped to a configuration where the scalar gradient
contributes anisotropically to the propagation of the wave.

Under the general conformal-disformal map \eqref{disformal2}, the same
qualitative picture persists: the pp-wave ansatz remains valid, with its
amplitude $\bar H$ rescaled by $A(X_0)$ and shifted by $B(X_0)$.  Thus, the
family of pp-wave metrics is kinematically stable under both conformal and
disformal redefinitions, although the dynamical decoupling between the scalar
and tensor sectors is generically lost.  Disformal transformations therefore act
as a bridge between stealth and nonstealth configurations within the broader
HOST framework.

Let us now consider the special case of a pure conformal transformation,
\begin{equation}
\label{conformal}
\bar g_{\mu\nu} = A(X)\, g_{\mu\nu},
\end{equation}
where $A(X)$ is an arbitrary function of the scalar kinetic term
$X = g^{\mu\nu}\phi_\mu\phi_\nu$.  For the pp-wave ansatz \eqref{pp_wave} with
the scalar configuration \eqref{scalar_pp}, the kinetic term takes the constant
value $X=X_0$.  Consequently, the conformal factor $A(X_0)$ is also constant,
and the transformation reduces to a global rescaling of the metric:
\begin{equation}
\bar g_{\mu\nu} = A_0\, g_{\mu\nu}, \qquad A_0 := A(X_0).
\end{equation}
The line element becomes
\begin{equation}
\label{pp_conformal}
\bar g_{\mu\nu} dx^\mu dx^\nu
= A_0 \bigl( H\, du^2 + 2\,du\,dv + dx_1^2 + dx_2^2 \bigr),
\end{equation}
which is manifestly conformal to the original pp-wave geometry.

Because $A_0$ is constant, the transformation does not distort the null
structure of the spacetime: the vector $\partial_v$ remains null, the surfaces
$u=\text{const}$ remain null hypersurfaces, and the causal structure is
unchanged.  In particular, the pp-wave condition
\begin{equation}
\partial_{x_1}^2 H + \partial_{x_2}^2 H = 0
\end{equation}
is preserved, since the Laplace operator on the transverse Euclidean plane is
invariant under constant rescalings.  The scalar kinetic term transforms as
\begin{equation}
\bar X = A_0^{-1} X_0,
\end{equation}
which remains constant and finite for $A_0>0$.  Even if $A(X)$ has a nontrivial
functional dependence on $X$, the constancy of $X$ on the pp-wave background
ensures that $A(X_0)$ acts as a constant.  Thus, the pure conformal map does not
introduce any new structure: it simply rescales all lengths uniformly.
This behavior stands in sharp contrast to the disformal case.  A disformal
transformation introduces an anisotropic deformation along the direction of
$\phi_\mu$, thereby modifying the transverse geometry and coupling the scalar
and tensor modes. 
By contrast, a pure conformal transformation preserves the isotropy of the $(x_1,x_2)$ plane and does not generate any mixing between the
scalar and gravitational sectors.
As a result, stealth pp-wave solutions remain
stealth under conformal transformations; hence, the scalar field continues to have no backreaction on the geometry, and the propagation of the gravitational wave is
unchanged up to an overall rescaling of its amplitude.
From a physical perspective, this reflects the fact that a constant conformal
rescaling does not alter the curvature tensors in any essential way.
The Weyl
tensor, which encodes the radiative degrees of freedom of the gravitational
field, is invariant under constant conformal transformations, and the Ricci
tensor rescales trivially.  Therefore, the conformally transformed metric
describes the same gravitational-wave solution, with the same profile $H(u,x_i)$
and the same decoupled scalar configuration.

To summarize, conformal and disformal transformations play distinct roles:

\begin{itemize}
  \item[(i)] A pure conformal map ($B=0$) preserves both the Laplace condition and
    the stealth nature of the pp-wave.
        
  \item[(ii)] A pure disformal map ($B\neq0$) preserves the Brinkmann form but not the
 Laplace condition, thereby coupling the scalar and metric perturbations
and destroying the stealth property.
\end{itemize}
Thus, while the conformal sector acts trivially on pp-wave geometries, the
disformal sector generates genuinely new nonstealth configurations within the
same HOST class.

\section{Conclusion}

In this work, we have presented a comprehensive analysis of exact plane-fronted
gravitational-wave (pp-wave) solutions in shift-symmetric quadratic-order
HOST theories, with particular emphasis on the
degenerate subclasses (DHOST).  By imposing only minimal geometric assumptions-
the existence of a null coordinate and a two-dimensional Euclidean transverse
space—we constructed the most general pp-wave ansatz obtainable after exhausting
the residual coordinate freedom.  Substituting this ansatz into the HOST action,
we showed that the Euler-Lagrange equations reduce to a single two-dimensional
Laplace equation for the metric profile $H(u,x_1,x_2)$, exactly as in vacuum GR.  This demonstrates that the mathematical structure governing pp-waves is remarkably stable under higher-order scalar-tensor extensions.

A key outcome of our analysis is that the pp-wave geometry and the scalar field
configuration are not independent: once the metric is required to take the
pp-wave form, consistency of the field equations forces the scalar field to be
linear in the transverse coordinates while remaining arbitrary in the retarded
time $u$.  This structure keeps the kinetic term $X$ constant, ensuring that the
scalar field remains dynamically active without modifying the spacetime
geometry.  The resulting algebraic conditions on the coupling functions are
mild, showing that stealth pp-wave solutions arise naturally within class–Ia
DHOST theories without fine-tuning or special symmetries.

We further examined the behavior of these solutions under conformal and
disformal transformations.  Pure conformal transformations act only as global
rescalings and preserve both the pp-wave condition and the stealth character of
the solution.  In contrast, disformal transformations preserve the Brinkmann
form of the metric but generically break the Laplace condition, thereby coupling
the scalar and tensor sectors and converting stealth configurations into
non-stealth ones.  This distinction highlights the subtle interplay between
scalar gradients and null wave propagation in DHOST theories.

Overall, our results show that pp-wave geometries provide a robust probe of the
structure of higher-order scalar-tensor gravity.  They reveal which aspects of
GR persist unchanged in DHOST theories, which features are modified, and how
scalar degrees of freedom can remain hidden or become dynamically relevant
depending on the frame.  Future work may explore the phenomenological
implications of these solutions for gravitational-wave observations, their
generalization to broader HOST models, and the systematic use of disformal
transformations as a solution-generating mechanism in scalar-tensor gravity.

Beyond these results, several promising directions for future research emerge.
A first natural step is to investigate the stability of the stealth pp-wave
solutions identified here.  While the background geometry is insensitive to the
scalar field, perturbations around the solution may reveal nontrivial mixing
between tensor and scalar modes. 
A systematic analysis of linear and nonlinear perturbations
would clarify whether stealth pp-waves can serve as viable strong-field
backgrounds in DHOST theories or whether they exhibit instabilities absent in
GR.

Another important direction concerns the observational implications of these
solutions.  Since pp-waves represent exact gravitational-wave geometries, they
provide a controlled setting in which to study how scalar degrees of freedom
propagate alongside gravitational radiation.  Disformal transformations, in
particular, generate configurations in which scalar and tensor waves are
intrinsically coupled.  Exploring how such couplings manifest in waveform
distortions, polarization content, or propagation speed could offer new avenues
for constraining scalar-tensor gravity with current and future gravitational-wave
detectors.

It would also be valuable to extend the present analysis to more general classes
of HOST theories, including models with broken shift symmetry, nontrivial scalar
potentials, or higher-order derivative interactions.  Whether the Laplace
structure of the pp-wave equation persists in these broader settings remains an
open question.  Similarly, generalizations to other exact wave geometries—such
as Kundt waves, gyratons, or impulsive wave solutions—may reveal additional
universal features shared between GR and higher-order scalar-tensor theories.

Finally, the role of disformal transformations as a solution-generating
mechanism deserves further exploration.  The fact that stealth pp-waves are
mapped to nonstealth configurations suggests that disformal maps can be used to
systematically construct new families of exact solutions with controlled scalar-
tensor interactions.  Developing a classification of pp-wave-like geometries
under the full conformal-disformal group may shed light on the deeper geometric
structure underlying DHOST theories and their relation to GR.

%%%%%%%%%%%%%%%%%
\section*{ACKNOWLEDGMENTS}
This work was supported by the funds of Butsuryo College of Osaka.

\appendix

\section{Geometric construction of the pp-wave-type metric}
\label{app_gauge}

In this appendix, we show explicitly how, under the assumptions stated in the main text, the metric can be brought to the form \eqref{pp_wave_general} by a suitable choice of coordinates, without using the field equations. We also clarify the geometric implications of further specializing this form to the canonical pp-wave metric, distinguishing it from the broader Kundt class.

We start from a metric of the form
\begin{align}
ds^2 = g_{uu}\,du^2 + 2 g_{uv}\,du\,dv + g_{vv}\,dv^2
      + 2 g_{ui}\,du\,dx^i + 2 g_{vi}\,dv\,dx^i + g_{ij}\,dx^i dx^j,
\label{app_general_metric}
\end{align}
where the components $g_{\mu\nu}(u,x_1,x_2)$ are independent of $v$, and
$i,j=1,2$ denote the transverse indices.  We further assume that the transverse
sector admits a two-dimensional Euclidean symmetry, in the sense that $g_{ij}$
can be brought to a conformally flat form by a suitable redefinition of
$(x_1,x_2)$.

\subsection{Removal of the $dv\,dx^i$ terms}

Consider the coordinate transformation 
\be
u' = u, v' = v + \alpha(u,x^i), 
x'^i = x^i. 
\ee
The coefficients transform as $g'_{v'i'} = g_{vi} + g_{vv}\,\partial_i \alpha$. If $g_{vv} \neq 0$, we can choose $\alpha$ such that $\partial_i \alpha = - g_{vi}/g_{vv}$, which locally eliminates the $dv\,dx^i$ terms. In the context of the pp-wave limit where $g_{vv} \to 0$, these terms must vanish for the vector $\ell^\mu$ to be orthogonal to the transverse hypersurfaces.

\subsection{Normalization of the $du\,dv$ term}

After eliminating $dv\,dx^i$, we perform a reparametrization $u \rightarrow u'(u)$. The coefficient $g_{uv}$ transforms as $g'_{u'v'} = (du/du') g_{uv}$. Provided $g_{uv} \neq 0$, we can set $g'_{u'v'} = 1$. As noted in the main text, if the resulting coordinate $v$ serves as an affine parameter along the null rays, $V=1$ is a natural geometric requirement consistent with Eq.~\eqref{KUNDT_COND}.

\subsection{Removal of the $du\,dx^i$ terms and gyratonic effects}

The $du\,dx^i$ terms can be addressed by transverse shifts $x'^i = x^i + b^i(u)$. The new components become $g'_{u'i'} = g_{ui} + g_{ij}\,\dot{b}^j$. For a nondegenerate $g_{ij}$, one can solve for $\dot{b}^j(u)$ to set $g'_{u'i'} = 0$.

It is important to emphasize that this removal is generally local. In a global context, if the transverse space has a nontrivial topology or if one considers the angular momentum of the wave source, these $g_{ui}$ terms may not be entirely removable and can encode gyratonic effects~\cite{Frolov:2005in, Podolsky:2014lpa}. In the present work, we restrict our attention to the case where such terms are globally absent, focusing on the standard pp-wave class without gyratonic contributions.

\subsection{Conformal flatness and the Brinkmann form}

In two dimensions, any Riemannian metric $g_{ij}$ is locally conformally flat, $g_{ij} dx^i dx^j = W(u,x)(dx_1^2 + dx_2^2)$. Combining all steps, we obtain the general form \eqref{pp_wave_general}.

Finally, we comment on the reduction to the canonical pp-wave form \eqref{pp_wave_host}. While the steps above are based on coordinate freedom, the additional requirements $I=0$ and $W=1$ are essential to satisfy the geometric conditions $\nabla_\nu \ell^\mu = 0$ and $\ell^\mu \ell_\mu = 0$ as defined in Eq.~\eqref{KUNDT_COND}. Without $I=0$, the vector $\ell^\mu$ would not be null;
without $W=1$, the null rays would not be parallel in the sense of a plane-fronted wave, and the metric would belong to the more general Kundt class. Thus, the specialization to Eq.~\eqref{pp_wave_host} is a choice of the specific geometric class of spacetimes being investigated in the HOST analysis.

\bibliography{refs}
\end{document}